\documentclass[12pt]{article}
\usepackage{authblk}
\usepackage{graphicx}
\usepackage{cite}
\usepackage{amssymb,amsmath}
\usepackage{multirow}
\usepackage{footnote}
\usepackage[para]{threeparttable}
\begin{document}
\title{Dynamics of K$^+$ counterions around DNA double helix in the external electric field: a molecular dynamics study}%
\author{O.O.~Zdorevskyi}
\author{S.M.~Perepelytsya}

\affil{Bogolyubov Institute for Theoretical Physics of the National Academy of Sciences of Ukraine, 14b, Metrolohichna Str., Kyiv 03143, Ukraine, zdorevskyi@bitp.kiev.ua}

\setcounter{page}{1}%
\maketitle

\begin{abstract}
The structure of DNA double helix is stabilized by metal counterions condensed to a diffuse layer around the macromolecule. The dynamics of counterions in real conditions is governed by the electric fields from DNA and other biological macromolecules. In the present work the molecular dynamics study {was} performed for the system of DNA double helix with neutralizing K$^+$ counterions and for the system of KCl salt solution in the external electric field of different strength (up to 32 mV/{\AA}). The analysis of ionic conductivities of these systems {has shown} that the counterions around the DNA double helix are slowed down compared with KCl salt solution. The calculated {values of ion mobility} are within (0.05$\div$0.4) mS/cm depending on the orientation of the external electric field relatively to the double helix. Under the electric field parallel to the macromolecule K$^+$ counterions move along the grooves of the double helix staying longer in the places with narrower minor groove. Under the electric field perpendicular to the macromolecule the dynamics of counterions is less affected by DNA atoms, and  starting with the electric field values about 30 mV/{\AA} the double helix undergoes a phase transition from double-stranded to single-strand state.
\end{abstract}

\section{Introduction}
\label{intro}
DNA macromolecule in a water solution is a strong polyelectrolyte inducing a condensation of small mobile positive ions (counterions) from the bulk to a diffuse layer around the double helix \cite{manning_1978,Frank_Kamenetski__1987}. The ionic cloud screens the negatively charged phosphate groups of DNA backbone reducing the electrostatic repulsion of the opposite strands of the double helix that is determinative for its stabilization \cite{Saenger}. In the cell nuclei the part of electrostatic charge of the DNA macromolecule is neutralized by the positively charged molecular groups of histone proteins, while the counterions screen the rest of the negative charge of the double helix \cite{Savelyev,Korolev2012}. {The structure and dynamics of the cloud of counterions are governed by strong electrostatic fields, produced by DNA and protein macromolecules. Therefore, the dynamical response of counterions on the external electrostatic field should be studied to understand the physical mechanisms of DNA-counterion interaction in the real systems.}

The counterions around the DNA double helix are localized within the layer of about 7 {\AA} thickness \cite{manning_1978}. The main properties of the counterion cloud are described within the framework of the mean field approximation using the Poisson-Boltzmann equation for the charged continuum around the uniformly charged cylinder as a DNA macromolecule \cite{manning_1978,Frank_Kamenetski__1987,Katchalsky,Obukhov}. These models {have shown} that the concentration of counterions is high near the DNA surface and gradually decreases with the distance that is in good agreement with the experimental data on small angle X-ray scattering (SAAX) \cite{Das,Andresen,Qiu}. {At the same time, the mean field approximation is limited to describe the localization of counterions in different regions of the DNA structure, inside the minor groove and major groove of the double helix, and near the phosphate groups of the macromolecule backbone. At the same time, the molecular dynamics simulations can give important information for understanding the organization of counterion cloud around the DNA macromolecule.}

{The simulation studies of the distribution of metallic counterions around the DNA double helix started from the very first research on atomistic modeling of nucleic acids \cite{2Clementi81,Singh755,Seibel6537,parallelism1985,gunsteren1986,nilges1987,Aatto1989}. The description of electrostatic interactions, in particular, the interactions with counterions, was one of the key problems that were effectively  adressed by the Ewald summation technique \cite{Aatto1989,Kollman1995}. After  long and hard efforts of many groups, a realistic description of DNA with counterions in atomistic simulations was given at the end of the last century (see the perspective in the reviews \cite{mocci2012,Aksimentiev2014}). The molecular dynamics simulations with the characteristic trajectory length of few nanosecond have revealed the main binding sites for Li$^+$, Na$^+$, K$^+$, Cs$^+$, and Mg$^{2+}$ and other counterions: the minor groove and the major groove of the double helix, the sites near the atoms of the phosphate groups \cite{Young1997,mackerell1997,Aatto98,Charge2000}. The localization of Na$^{+}$ ions in the minor groove of the double helix contacting directly with the atoms of nucleotides has been established \cite{Young1997}, explaining the experimentally observed localization of counterion (Cs$^+$ ion) in the spine of hydration in the minor groove of the double helix \cite{bartenev}. The effects of counterion hydration in the case of different metal ions with different character of hydration have been found important for the interaction of counterions with DNA \cite{mackerell1997,Aatto98}.}

{The simulations with longer simulation trajectories (several tens nanoseconds) have established the characteristic residence times of metal counterions (Na$^+$ and K$^+$ mostly) and the  dependence of their positioning on the sequence of nucleotide bases \cite{varnai2004,Ponomarev2004,rueda2004,Mocci}. The results have shown that the counterions may stay in the binding site during the period from tens picoseconds up to one nanosecond \cite{varnai2004}. In the minor groove of the double helix the residence time is the longest, while in the major groove and near the phosphate groups the residence time is several times lower than in the minor groove \cite{varnai2004,Ponomarev2004,Mocci}. The atomistic simulations with much longer trajectories (about 1 $\mu$s) have proved these results and revealed new important physical features of the system \cite{Aksimentiev2012,Lavery2014,Maddocks,dans2016,Canadian}. In particular, the microsecond simulations \cite{Lavery2014,Maddocks} have shown that the systems of DNA with counterions come to equilibrium very slowly, and it takes about 100 ns or even more for the system to reach balance. Thus, on the basis of the results of atomistic molecular dynamics simulations \cite{varnai2004,Ponomarev2004,rueda2004,Mocci,Aksimentiev2012,Lavery2014,Maddocks,dans2016,Canadian}} the counterions may be classified by the region of their localization with respect to the double helix: in the solution ({the} free ions), in the cloud outside the macromolecule ({the} cloud ions), and inside the  minor and major grooves of the double helix ({the} bonded ions). The response  of counterions on the action of the electric field is expected to be different in the case of different  compartments of the double helix.

The dynamics of the ions in water solutions may be characterized by the ionic current that appears under the action of the external electric filed. The experimental studies for DNA in water solution with the different salts of alkali metal ions show that the ionic conductivity essentially depends on the temperature, counterion type and concentration \cite{Grassi,Kranck,Kuznetsov1981,Kuznetsov1987,Vuletic_2011,Liubysh2014}. The dependence of conductivity on counterion type is {featured} by the ion {size, }charge and character of hydration \cite{Izmailov}. The temperature dependence of the conductivity is related to the increase of the ion mobility with the temperature increase and with the conformational transformations of the DNA occurring due to the melting of the double helix \cite{Kuznetsov1981,Kuznetsov1987}. The study of concentration dependence of ionic conductivity for DNA with KCl salt {has shown} that at salt concentrations lower than 0.4~M the conductivity of DNA solution is higher than the conductivity of KCl water solution without DNA \cite{Liubysh2014}. This effect {has been} explained as the result of ordering of counterions around the DNA double helix in a lattice-like structure, the existence of which was studied by the specific ion-phosphate vibrations in the DNA low-frequency spectra $<$200 cm$^{-1}$ \cite{PV_EPJE_2007,PV_EPJE_2010}. Thus, the experiments on the DNA conductivity in solutions can give an information about average characteristics of the dynamics of the counterions.

To elucidate the microscopic {structure and} dynamics of DNA {with the} counterions, hopped by the external electric field, the molecular dynamics simulations have been carried out for the case of different ion types \cite{Aksimentiev2014,kowalczyk_2012,comer_2012,belkin2016,Xiang_2018}. In particular, the counterion dynamics during the motion of the DNA macromolecule through the nanopore was in the focus of the interest due to the  potential application of the measuring of electric current in the DNA sequencing technologies \cite{kowalczyk_2012,Xiang_2018}. To improve the nanopore experiments and make them more controllable and predictable the self-assembling DNA origami structures ~\cite{DNAorigami} were used. The conductivity values {have been} shown to be very sensitive to the lattice type of DNA origami as well as to the concentration of the counterions in solution~\cite{aksimentiev2015}. {The study of the dynamics of K$^+$ and Li$^+$ counterions has shown that in the immediate vicinity of the DNA surface, the ion mobility is about a half lower than in solution, and inside the macromolecule, it is even less \cite{comer_2012,belkin2016}. This fact reveals the strong influence of the DNA macromolecule on the dynamical properties of ions. The} modelling study, comparing the dynamical response of the ions on the presence of the external  electric field in the solution with DNA and without DNA, can give important insights to the behaviour of the counterion cloud around DNA macromolecule in real conditions.

The goal of the present research {was} to compare the dynamical response of K$^+$ ions around the DNA double helix and in water solution without DNA macromolecule on the action of the external electric filed. To solve this problem the molecular dynamics simulations have been carried out for the K-DNA and for the solution of KCl in the external electric field of different strength. The cases of parallel and perpendicular directions of the electric field with respect to the helical axis of the DNA macromolecule {have been} considered. The mobility of the counterions{,} localized in different compartments around the double helix has been analysed. The results {have shown} that the dynamics of counterions outside the double helix is resemble to the dynamics in the solution, while in the DNA grooves the motion of the counterions is modulated by the structure of the double helix.

\section{Materials and methods}
\label{sec:methods}

\subsection{Simulation setup}
\begin{figure}
\begin{center}
\resizebox{0.7\textwidth}{!}{%
  \includegraphics{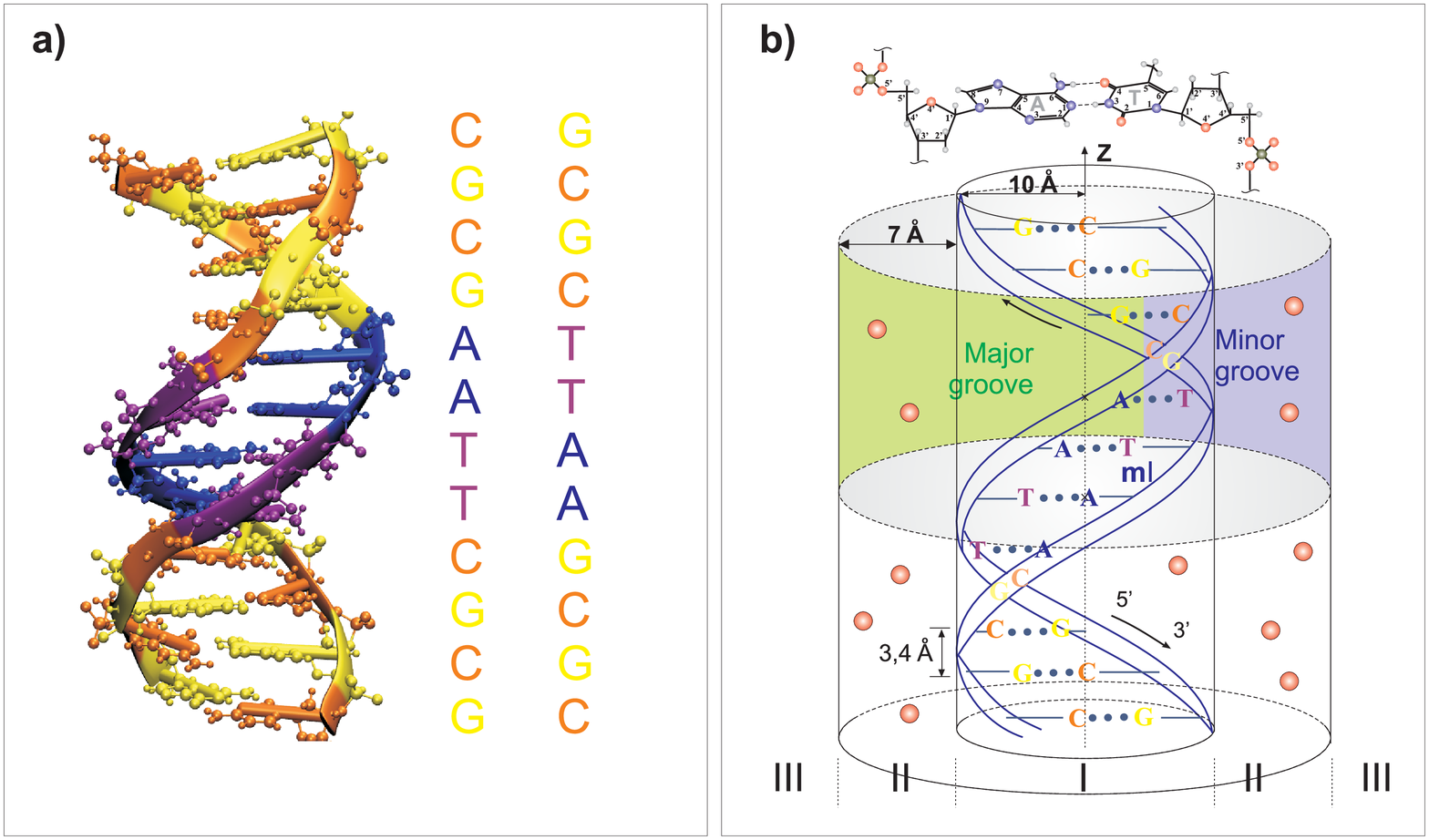}
  }
\caption{a) The structure of DNA fragment in K-DNA system -- Drew-Dickerson dodecamer d(CGCGAATTCGCG). The nucleotide colour scheme: Cytosine (orange), Guanine (yellow), Adenine (blue), Thymine (purple). b) Compartments of the DNA double helix. The minor and major grooves are coloured as yellow and gray, respectively. The first (I), second (II), and third (III) hydration shells are indicated.}
\label{fig:systems1}       % Give a unique label
\end{center}
\end{figure}

The atomistic molecular dynamics simulations of two molecular systems in the electric field have been carried out. The first system was the fragment of DNA double helix with the nucleotide sequence d(CGCGAATTCGCG) amerced into the water box with the size 64$\times$64$\times$64 {\AA}$^3$ (K-DNA system). This polynucleotide, known as Drew-Dickerson dodecamer ~\cite{1bna}, is commonly used in the molecular dynamics simulations of DNA systems \cite{mackerell1997,Charge2000,rueda2004,Ponomarev2004,mocci2012}. In the case of the considered dodecamer the minor groove of the double helix is narrowed in the AATT nucleotide region compared to the average value, while the major groove is visibly wider (Fig. \ref{fig:systems1}). The system was neutralized by 22 K$^+$ counterions. The number of positively charged counterions was equal to the number of negatively charged phosphate groups of DNA. The second system was the water solution of KCl salt with the concentration 139 mM (KCl system). Such concentration corresponds to 22 pairs of K$^+$ and Cl$^-$ ions. The number of water molecules was 7917 and 7977 in K-DNA and KCl systems, respectively.

The computer simulations {were} performed using NAMD software package \cite{Phillips} and CHARMM27 force field \cite{mackerell_2000,foloppe_2000}. The VMD software package {was} used for the systems construction, visualization and analysis~\cite{VMD}. The structural parameters of the DNA double helix have been calculated using the 3DNA~\cite{3dna} {and do\_x3dna~\cite{kumar2015do_x3dna} analysis tools}. The lengths of all bonds with hydrogen atoms were taken rigid using SHAKE algorithm \cite{SHAKE}. The TIP3P water model \cite{TIP3P} and the Beglov and Roux parameters of ions  \cite{beglov_1994} {were} used. The Langevin dynamics {was} used for all heavy atoms with the temperature 300 K. The long-range electrostatic interactions {were} treated using particle mesh Ewald method \cite{PME}. The switching and cutoff distances for the long-range interactions {were} 8 {\AA} and 10 {\AA}, respectively. The periodic boundary conditions {were} used.

{The simulation protocol was composed on the basis of experience of the previous molecular dynamics studies of the dynamics of DNA with counterions \cite{Young1997,Aatto98,varnai2004,Ponomarev2004,rueda2004,Mocci}.} K-DNA system was initially simulated at the constant pressure 101325 Pa and temperature 300 $^{\circ}$K (NPT ensemble). The procedures of minimization, heating and equilibration were performed for the system with restrained DNA atoms. The restraints were made gradually weaker, and after about 2 ns of equilibration all DNA atoms were free to move, except the atoms of C-G nucleotide pairs in the ends of the double helix that were restrained by the force with the coefficient k=5 kcal$/$mol{\AA}$^2$. Then the system was equilibrated in NVT ensemble during 200 ns with the restrained ends of DNA. In the case of KCl system the simulation stages in NPT ensemble were performed without any restrains, and the equilibration in NVT ensemble was for 50 ns.

{Harmonic restraints applied to the ends of the DNA fragment were used to avoid a drift of the macromolecule and its rotation. The trajectory of the center of mass motion and the change of orientation of the double helix in the case of the absence of the restrains and without external electric field are demonstrated in Fig. S1 (Suppl. mat.). The results have shown that even without the external electric field the motion of the DNA macromolecule as a whole and its rotation are significant. In the reference frame related with the double helix, the rotation of the macromolecule causes change in the direction of the external electric field, which is undesirable in the present research. Therefore, to study the action of external electric field the simulations have been performed with restrained end nucleotide pairs of the double helix.}

{Our study of the influence of the restraints on the structure of macromolecule has shown changes in the secondary structure of DNA. In particular, the widths of the minor and major grooves of the double helix have been analyzed (Suppl. mat., Fig. S2). The results have shown that the dependence of the major groove width on the sequence of nucleotide bases has the same character in the case of DNA with the restrained end nucleotides and without restrains. The width of the minor groove in the case of restrained ends of the macromolecule differs essentially only in the case of the boundary pairs. However, the structural changes of the DNA double helix induced by restrains are not strong enough to change the physical conclusions of the present research.}

\begin{figure}
\begin{center}
\resizebox{0.6\columnwidth}{!}{%
  \includegraphics{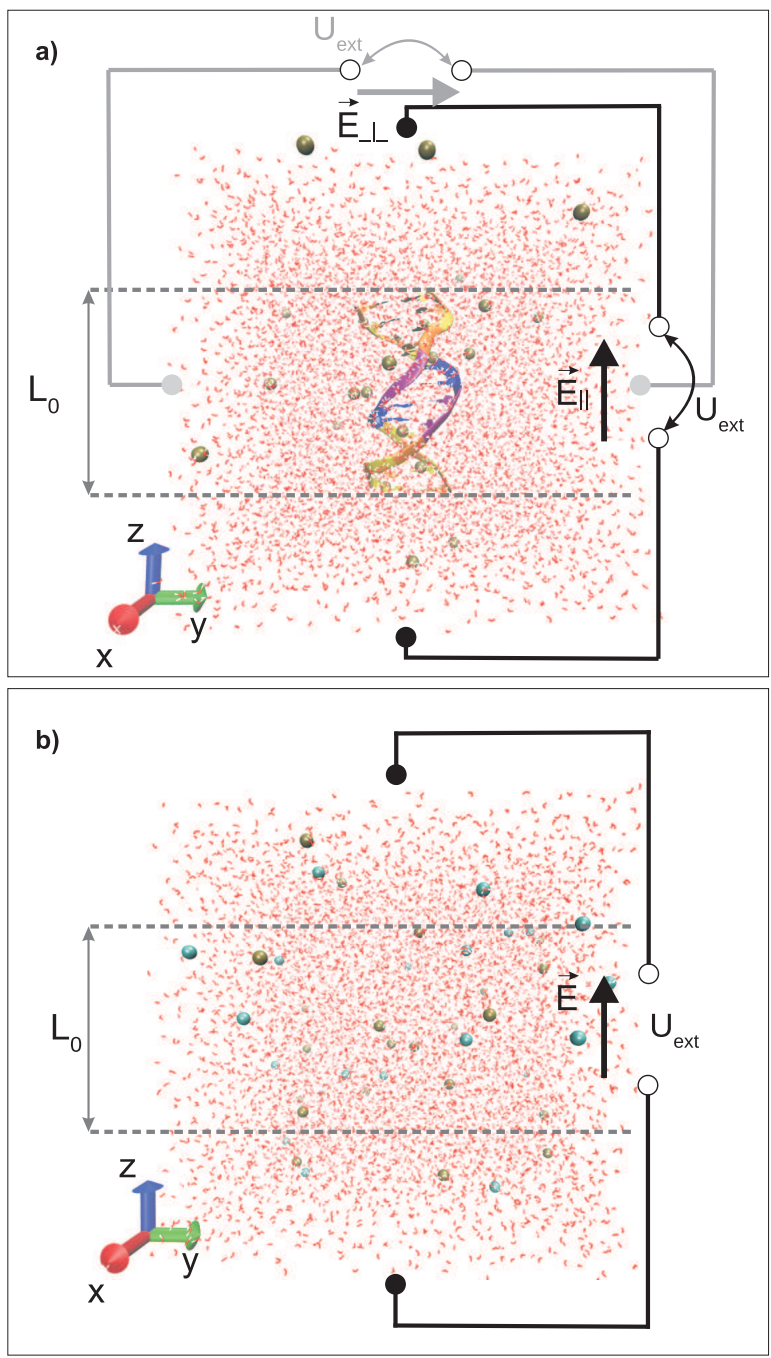}
  }
\caption{a) Scheme of the K-DNA system. The external electric field is applied to the system in the direction of z axis (E$_{\parallel}$) and of x axis (E$_{\perp}$). b) Scheme of KCl system: K$^+$ (brown) and Cl$^-$ (blue) ions in water solution. For the both systems, the selection of ions situated only between the two dashed lines are taken into account. L$_0$=37.4{\AA} corresponds to the length of the DNA fragment.}
\label{fig:systems2}       % Give a unique label
\end{center}
\end{figure}

After the equilibration stage the systems were simulated in the external electric field, produced by applying the electric potential to the boundaries of each system (Fig. \ref{fig:systems2}). In the case of KCl system only one orientation of the electric field was necessary to consider (Fig \ref{fig:systems2} a). For K-DNA system the cases of parallel (along $z$ axis) and perpendicular (along $x$ axis) orientations of the electric field to the helical axis were considered (Fig. \ref{fig:systems2} b). The systems were placed in the electrostatic field with the voltage values 145, 290, 580, 870, 1450, and 2030 mV. Such voltages are of the same order of magnitude as electrical potential at the distance 10 {\AA} from the point elementary charge in water medium with dielectric constant $\varepsilon$=80. This value also corresponds to the electric field that was used for the simulations of ionic currents in the systems with DNA origami ~\cite{aksimentiev2015}. Each system has been simulated during 50 ns after equilibration stage for the applied electrostatic filed with the defined direction and voltage value.

\subsection{Analysis of the trajectories}

The obtained simulation trajectories for K-DNA and KCl systems have been analyzed to characterize the dynamical response of the ions on the presence of the electric field. In particular, the electrical conductivity caused by the ions of the systems (ion conductivity) has been calculated. The contributions to the ion conductivity from the counterions, localized in different compartments of the DNA double helix, have been derived. %The analysis of the simulation trajectories has been carried out using the VMD software package \cite{VMD}.

The current density {j} induced by the electric field {E} is governed by  Ohm's law:
\begin{equation}\label{Eq1}
\mathbf{j}=\sigma \mathbf{E},
\end{equation}
where $\sigma$ is the electrical conductivity of the system. In terms of the system current $I$ and the external bias $U$ the formula (\ref{Eq1}) may be written in the following form:
\begin{equation}\label{Eq2}
I=\frac{{\sigma}S}{d}U,
\end{equation}
where $d$ is the distance between the boundaries of the system to which the bias is applied, and $S$ is the area of the surface perpendicular of the external electric field.

{The method for the calculations of ionic current through the atomistic molecular dynamics simulations has been developed in~\cite{aksimentiev2015}, and in the present research we follow this method in all details.} Let us consider the rectangular box of the volume $V=L_xL_yL_z$, placed in the electric field $\mathbf{E}$ directed along $z$ axis. The ionic current may be defined in the following form:
\begin{equation}\label{Eq3}
I=\sum_{i}{\frac{\Delta Q_i}{\Delta t}},
\end{equation}
where $\Delta Q_i$ is the partial charge, transmitted by the $i$-th ion through the surface area $S$ by the time $\Delta t$. Taking into consideration that the $i$-th ion shifts with the time $\Delta t$ for the distance $\Delta z_i$, the partial charge may be determined as $\Delta Q_i=q_i\Delta z_iL_zL_y/V=q_i\Delta z_i/L_z$, where $q_{i}$ is the charge of the $i$-th ion. As the result, the electric current that flows  through the simulation box at the moment $t$ may be calculated by the following formula {\cite{aksimentiev2015}}:
\begin{equation}\label{Eq4}
I\left(t+\frac{\Delta t} 2\right)=\frac{1}{\Delta tL_z}\sum_i{q_i\Delta z_i}.
\end{equation}
Here, the current value is related to the time at the half step $t+\Delta t/2$. {To deal with periodic boundary conditions in the calculations of the displacements $\Delta z_i$, the following rules have been taken into consideration~\cite[S2]{aksimentiev2015}}:
\begin{equation}\label{Eq5}
\Delta z_i=
\begin{cases}
\Delta z_i(t) \;\ \texttt{if} \;\ |\Delta z_i(t)|<L_z/2; \\
\Delta z_i(t)-L_z \;\ \texttt{if} \;\ \Delta z_i(t)>L_z/2; \\
\Delta z_i(t)+L_z \;\ \texttt{if} \;\ \Delta z_i(t)>-L_z/2;
\end{cases}
\end{equation}
where $\Delta z_i(t)=z_i(t+\Delta t)-z_i(t)$.
Using the formulae (\ref{Eq4}) and (\ref{Eq5}) the ionic current has been calculated. The obtained current values are averaged over all simulation trajectory.

{The calculated dependencies of ionic current upon time have shown (Fig. S3, Suppl. mat.) that for all the considered external voltage values, the ionic current reaches a plateau within the first nanosecond of the simulation. After that, the ionic current fluctuates  around some average value. Thus, it may be concluded that the length of the each our simulation (50 ns) is enough to analyze the stationary process of current flow. }

To {analyze} different contributions to the electric current the space around the DNA double helix has been split into the following compartments: the first (I) the second (II) and the third (III) hydration shells of the macromolecule (Fig. \ref{fig:systems1} b). The ions were considered localized in the first hydration shell if they were within 10 {\AA} of atoms N$_1$ of Adenine or Guanine. The ions belonging to the second hydration shell were out of the first hydration shell but within 7 {\AA} of any atom of the DNA structure. The third hydration shell is supposed to be all the remaining space outside of the I and II hydration shells. To reduce boundary effects only those ions that have the same z coordinates as the atoms of double helix were taken into account (the space between the two dashed lines in Fig. \ref{fig:systems2}).

Analysing the contributions to the conductivity that make the ions from different compartments of the double helix, the fact that  the  number of ions in some region fluctuates with the time should be taken into consideration. In this regard,  the dynamics of the ions should be characterized by some physical parameter that is related to every single ion. As a convenient parameter the ion mobility ($\lambda$) is taken. In the case of our simulations this parameter is calculated as follows:
\begin{equation}\label{Eq6}
\lambda=\frac{1}{n_f}\sum_{f=1}^{n_f}\frac{\sigma_f}{N_f},
\end{equation}
where $n_f$ is the number of frames in the simulation trajectory; $\sigma_f$ is the conductivity, related to some compartment of the system, at the frame $f$; $N_f$ is the number of counterions in the compartment at the frame $f$.

The distribution of counterions around the DNA double helix {were} calculated for the first and the second hydration shells of the double helix. To characterize  ionic distributions  the number of ions has been calculated in the segments related to each nucleotide pair. The center of each segment {was} taken as the center of mass of the nucleotide pair. Since the distance between the adjacent base pairs equals to 3.4 {\AA}~\cite{Saenger}, the boundaries of each segment are located at the distance $\pm$1.7{\AA} along the $z$ axis from its mass center. {Actually, during the simulation, under the action of the external electric field, the structure of double helix can be slightly deformed, i.e. the center of each segment will not correspond exactly to center of mass location of the certain base pair. The performed analysis has shown that in the case of parallel electric field the curvature of double helical axis is $\sim$ 0.12 {\AA}$^{-1}$ (Fig. S4, blue line, Suppl. mat.). In the case of perpendicular electric field, the curvature values are not much higher, but starting from the bias values greater than $\approx$ 1000 mV, the curvature of DNA fragment increases essentially (Fig. S4, red line, Suppl. mat.). Thus, the external electric field influences the curvature of the double helix, but this effect is not so significant to change the conclusions of our work.}
\section{Results}

\begin{figure}[h]
\begin{center}
\resizebox{0.5\columnwidth}{!}{%
  \includegraphics{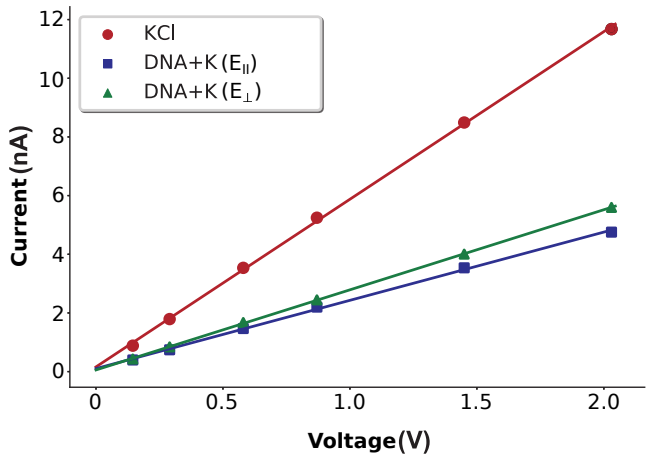}
  }
\caption{The dependence of ion current on a voltage in the simulated systems: KCl system (red circles), K-DNA system with the electric field $E_\parallel$ parallel to the helical axis  (blue squares) and with the electric field $E_\perp$ perpendicular to the helical axis (green triangles). The lines show linear approximation of the calculated data.}
\label{fig:KDNAKCl}       % Give a unique label
\end{center}
\end{figure}

\begin{figure}[h]
\begin{center}
\resizebox{0.5\columnwidth}{!}{%
  \includegraphics{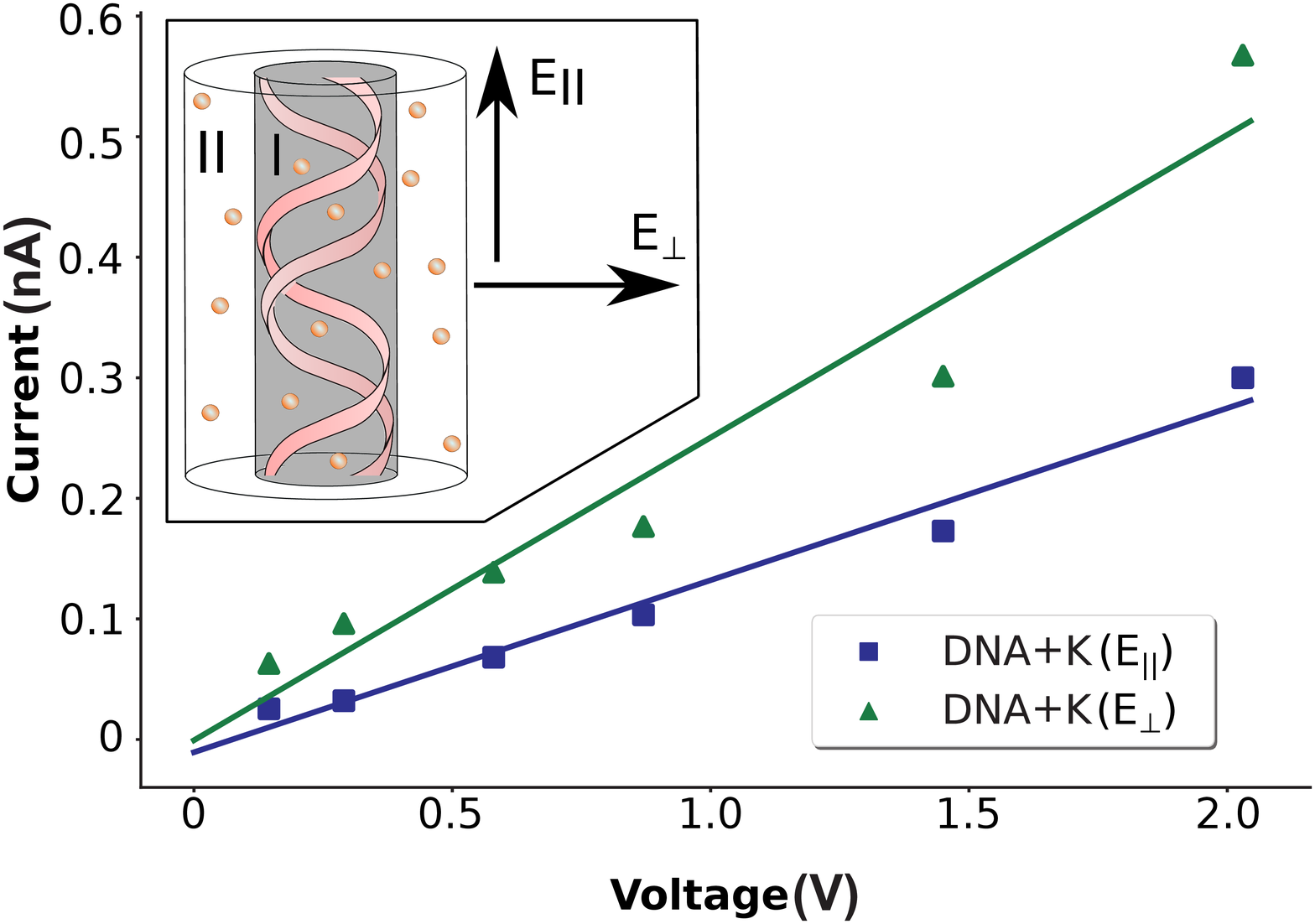}
  }
\caption{The calculated in the present work volt-ampere dependence for the first hydration shell of the DNA double helix (K-DNA system). The points, blue square and green triangles, show the results for the case of parallel ($E_\parallel$) and perpendicular ($E_\perp$) orientations of the electric field to the helical axis, respectively. The lines show the linear approximation of the calculated data. In the inset the first hydrations shell of the DNA double helix is schematically shown.}
\label{fig:VoltAmperInside}       % Give a unique label
\end{center}
\end{figure}

Using the obtained simulation trajectories the ionic current has been calculated for different voltage values. As the result, the volt-ampere characteristics have been obtained for K-DNA and KCl systems (Fig. \ref{fig:KDNAKCl}). The electric current increases as the voltage increases. The dependencies are almost linear that confirms the validity of the Ohm's law for the considered systems. The ionic current in the KCl solution increases much faster than in the case of K-DNA system that is due to the twice higher number of charge carriers in the KCl system (the ions K$^+$ and Cl$^-$) than in K-DNA system (the ions K$^+$). The difference of volt-ampere characteristics for the case of parallel and perpendicular orientations of the electric field with respect to the axis of the DNA double helix {has been} observed. The analysis of the currents in different compartments of the K-DNA system {has shown} that in the first hydration shell of the double helix the current is about an order of magnitude lower than the current of the whole system (Fig. \ref{fig:VoltAmperInside}). In the case of the parallel orientation of the electric field the volt-ampere characteristics is almost linear, while in the case of the perpendicular orientation of the electric field it increases faster. This effect appears due to the interaction of the counterions with the DNA double helix.

The slopes of the lines of the volt-ampere dependencies determines the ionic conductivity of the systems $\sigma$. The calculated values of the conductivities of the systems are presented in the Table 1. The results {have shown} that the highest conductivity is in the case of KCl system. The presence of a DNA macromolecule in the solution reduces the conductivity of the system, and in the case of the parallel orientation of the electric field the influence of DNA is stronger. The sum of the contributions from the K$^+$ and Cl$^{-}$ ions to the total conductivity of the KCl solution results the total conductivity of the  KCl system. The same is in the case of K-DNA system: the total conductivity of the system is a sum of contributions from the ions of different compartments of the double helix (the first, the second and the third hydration shells). Thus, the conductivity of the simulated KCl and K-DNA systems is the additive value of the constituents of the systems.

\begin{table}
\begin{center}
\caption{The conductivity $\sigma$ and mobility $\lambda$ of the ions in K-DNA and KCl systems.}
\label{tab:results}       % Give a unique label
\begin{tabular}{llcc}
\noalign{\smallskip}\hline\noalign{\smallskip}
	system	       &	component&	$\sigma$ (mS/cm)		& $\lambda$ (mS/cm)	\\[2mm]
\noalign{\smallskip}\hline\noalign{\smallskip}
K-DNA              &             &              			&	                        \\[2mm]
$E_\parallel$      & I           &	0.22		           	&	0.05	            	\\
	 	           & II          &    1.16               	&	0.29		            \\
	 		       & III         &		2.26	           	& 0.40                     	\\
	           	   & total       &		 3.64          		&0.26                     	\\[2mm]
$E_\perp$	 	   & I           &		0.39	         	&0.11	                    \\
		           & II 	     &		0.81	        	&0.25                      	\\
	 		       & III         &		3.09		        & 0.34                    	\\
                   & total  	 &		4.28		        & 0.27                  	\\[2mm]
KCl	               &             &		            		&			                \\[2mm]
		           & K$^+$    	 &		4.56		        &  0.36                     \\
 	               & Cl$^-$	      &		4.39		        &	0.34                    \\
                   & K$^+$+Cl$^-$ &		8.95     	     	&0.35                     	\\[2mm]
\noalign{\smallskip}\hline
\end{tabular}
\end{center}
\end{table}

\begin{figure*}
\begin{center}
\resizebox{1.0\textwidth}{!}{%
  \includegraphics{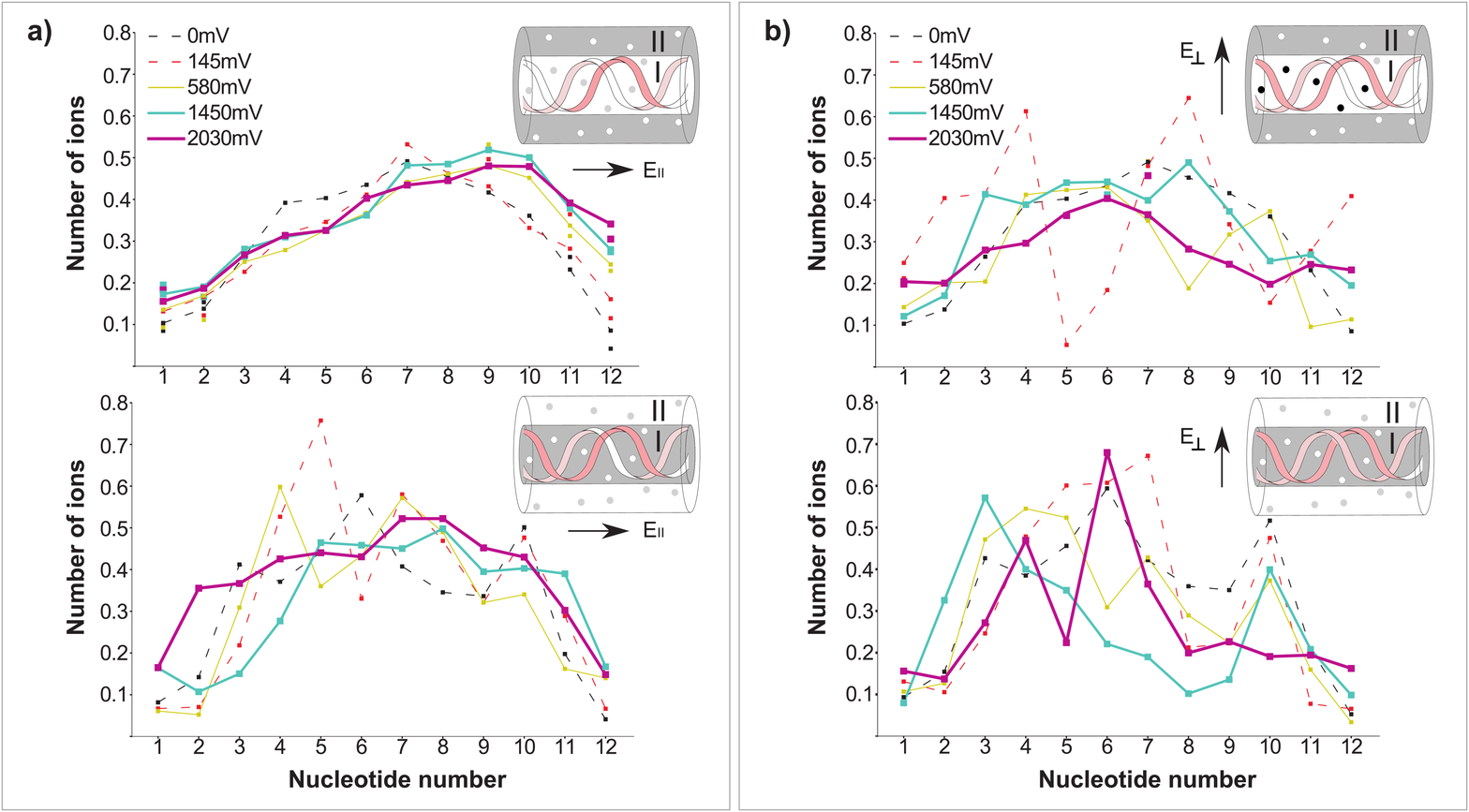}
  }
\end{center}
\caption{Distribution of ions along the double helix in the K-DNA system for the different values of the parallel (a) and perpendicular (b) orientations of the external electric field. The top graphs take into account ions from the second hydration shell (coloured in grey in the inset) and the bottom graphs - ions from the first hydration shell. These results are averaged over the last 10 nsec of the each 50-nsec simulation.}
\label{fig:ionsDistrib}       % Give a unique label
\end{figure*}
\begin{figure}
\begin{center}
\resizebox{0.6\columnwidth}{!}{%
  \includegraphics{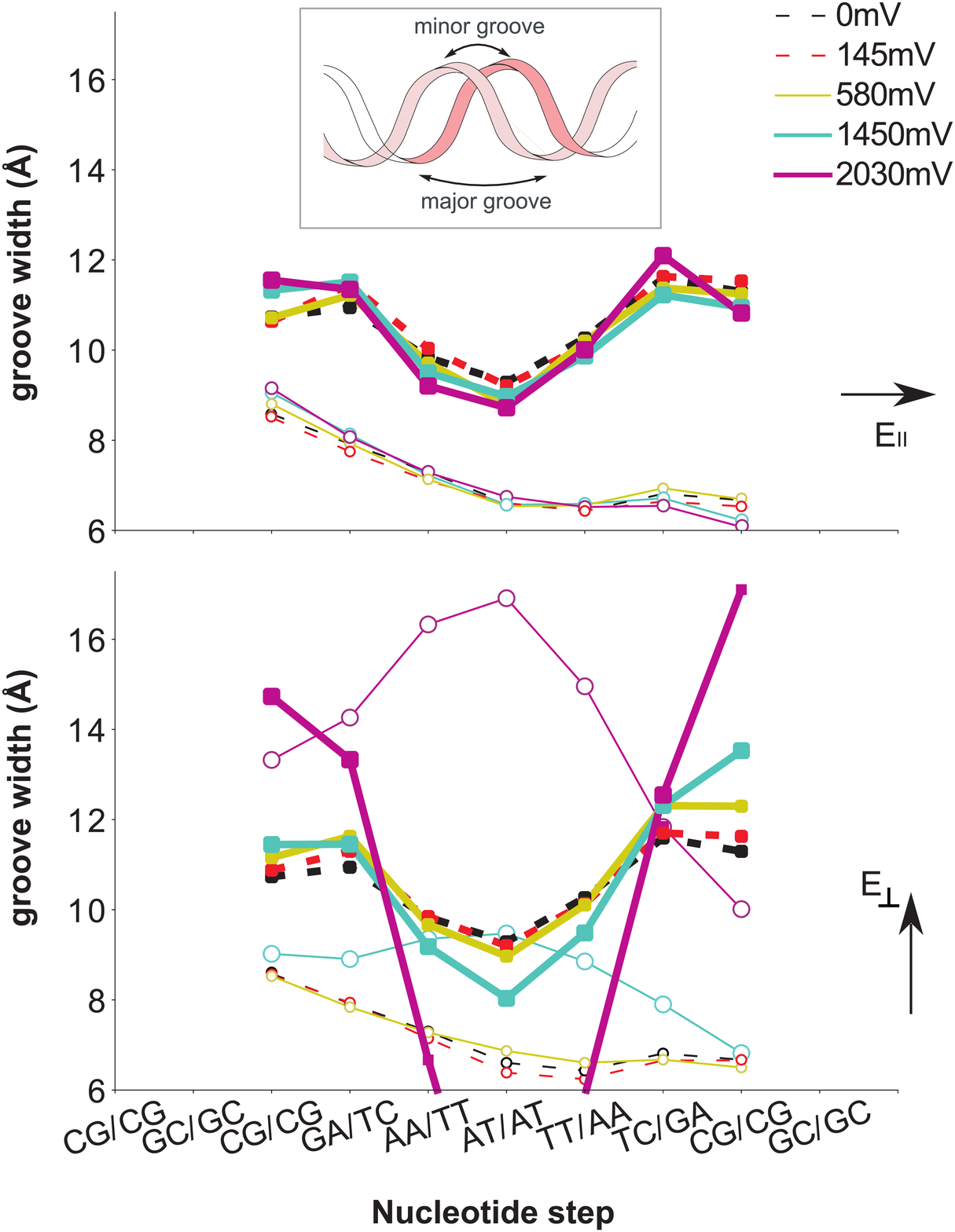}
  }
\caption{Widths of the minor (thin lines with circles) and the major (thick lines with squares) grooves of the DNA double helix upon the nucleotide steps in the case of the parallel (top graph) and perpendicular (bottom graph) orientations of the external electric field to the macromolecule. {Standard deviations are shown in Fig. S5 (Suppl. mat.).}}
\label{fig:grooves}       % Give a unique label
\end{center}
\end{figure}

The calculations of the mobility values {have revealed} that the mobility of K$^+$ and Cl$^-$ ions in KCl system are almost equal (around 0.35 mS/cm). In the K-DNA system the highest mobility is in the case of the ions that belong to the third hydration shell of the macromolecule. In the case of the parallel orientation of the electric field to the helical axis this value (around 0.4 mS/cm) is even higher than the mobility in the bulk KCl solution. Such difference may appear due to the friction of fluxes of K$^+$ and Cl$^-$ ions present in KCl system but not in the K-DNA system. {The mobility of the counterions in the second hydration shell of the macromolecule is characterized by the values lower than in the third hydration shell, and in the case of the first hydration shell, it is even lower. Given this, the following order of ionic mobilities (Tabl. \ref{tab:results}) may be presented: $\lambda_I<\lambda_{II}<\lambda_{III}$, where the indices designate the hydration shell of DNA (Fig. \ref{fig:systems1} b). This result qualitatively reproduces the dependence obtained in \cite{belkin2016} for the mobility of K$^+$ counterions distanced from the DNA double helix localized in nanopore.}

%The paragraph is split into two paragraphs

{The obtained mobility values of K$^+$ counterions in the third and second hydration shells} are almost the same in the case of the parallel and perpendicular orientations of the external electric field with respect to the double helix. The mobility values of the ions in the first  hydration shell of DNA under the parallel orientation of the electric field to the macromolecule is more than twice lower than the corresponding value for the perpendicular orientation of the electric field (Table 1). This effect can be explained by the fact that the counterions, hopped by the electric field parallel to the macromolecule, pass much longer distance near the DNA surface. Note, the mobility values of the ions, averaged over all hydration shells of the macromolecule are almost the same (about 0.26 mS/cm) in the case of $E_\parallel$ and  $E_\perp$ that is due to the relatively low volume occupied by the DNA fragment comparing to the total volume of the simulations box.

The analysis of the distribution of  counterions along the DNA double helix {has shown} that the number of ions in the first and second hydration shells of the macromolecule depends on the sequence of the nucleotide bases. The calculated dependence of the number of ions on the nucleotide sequence has a maximum in the vicinity of the center of the DNA fragment and goes down in the ends of the dodecamer (Fig. \ref{fig:ionsDistrib}). There are two reasons of such non-uniform behaviour. The first reason is the decrease of the electrostatic field strength in the end of the macromolecule (end effects). The second reason is related to the structural effects, in particular, the presence of AT-rich sequence  in the central part of Drew-Dickerson dodecamer, as the minor groove in this region is narrower.

The results of our molecular dynamics simulations {have shown} that the width of the minor and major grooves depends on the external electric field. Under the natural conditions the differences in the width of the minor groove from the ends to the center of the fragment can be up to 2 {\AA}, while the major groove is characterized by the narrow part at the center and wider parts at the ends of the simulated DNA macromolecule (Fig. \ref{fig:grooves}). Under the presence of the external electric field directed parallel to the helical axis, $E_\parallel$, the increase of the field strength makes the minor groove wider at one end of the DNA fragment and narrower at the other (Fig. \ref{fig:grooves}, top graph, thin lines with circles). The major groove of the DNA double helix also undergoes changes: in the central part the width of the major groove becomes narrower, while in the end it becomes wider under the influence of the electric filed $E_\parallel$ (Fig. \ref{fig:grooves}, top graph, thick lines with squares).

The structural response of the double helix on the presence of the external electric field parallel to the DNA double helix results the shifting of the maxima of the ionic distribution for two nucleotide pairs in the direction of the external electric field (Fig. \ref{fig:ionsDistrib} a, top graph). The correlation between width of the minor groove and the position of the maximum of the number of counterions is observed. The external electric field makes the distribution of ions much smoother, since in this case  the Coulomb forces  acting on a counterion from the external electric field is stronger than the attraction forces from the atomic groups of the DNA double helix. The number of ions localized in the regions of boundary nucleotide pairs (area from 8th to 12th nucleotide) increases with the applied voltage. This effect can be explained by the fact that under the action of higher external forces and consequently the extension of the minor groove ions start to abandon their {\textquoteleft}residence places{\textquoteright} much more often.

In the case of the perpendicular orientation of the electric field, $E_\perp$, the structural changes are not significant at the voltage values lower than 1450 mV and the dependence of the width of the minor and major grooves are qualitatively the same as in the case of the absence of the external electric filed (Fig. \ref{fig:grooves}, bottom graph). Starting with the voltage value 1450 mV the width of the minor groove in the center of the DNA fragment becomes maximal, and the width of major groove drastically decreases (Fig. \ref{fig:grooves}, bottom graph, blue and purple lines). Due to the significant increase of the minor groove width the opening of the nucleotide pairs in the central part of the macromolecular fragment {has been} observed, and the {\textquoteleft}breakdown{\textquoteright} of the DNA double helix takes place. The structural changes of the double helix also influence the ionic distribution: the average number of the ions in the first hydration shell of the double helix firstly decreases with the increase of the applied bias, but then it starts to increase (Fig. \ref{fig:ionsDistrib} b, bottom graph). This effect {has been} also observed in the volt-ampere characteristics that becomes non-linear (Fig. \ref{fig:VoltAmperInside}, green triangles). {Note, that the widths of the major and minor grooves experience significant fluctuations that are $\sim$ 1 {\AA} under E$_\parallel$ and up to $\approx$ 2.5 {\AA} in the case of E$_\perp$ at the central part of the DNA fragment (Fig. S5, Supp. mat.).}

\section{Discussion}

\begin{figure}
\begin{center}
\resizebox{0.6\columnwidth}{!}{%
  \includegraphics{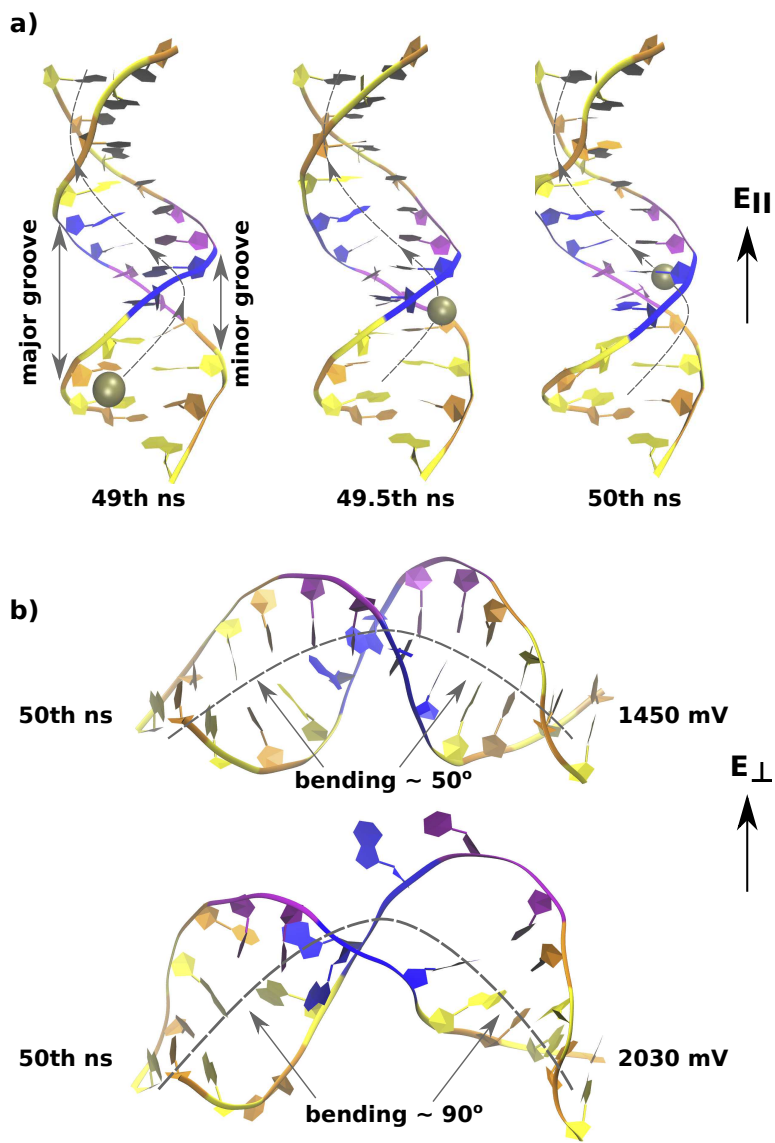}
  }
\caption{a) Snapshots from the the last nanosecond of the simulation of the K-DNA system with $E_\parallel$ and the external voltage of 2030 mV. The timestep between each of the adjacent snapshots is 0.5 ns. To simplify the visualization only the one ion that is moving along the DNA minor groove is shown; b) Visualization of the conformational transition of DNA under the influence of $E_{\perp}$. Snapshots are taken from the last frame of the simulations with the external biases of 1450 mV (top) and 2030 mV (bottom).}
\label{fig:snapshots}       % Give a unique label
\end{center}
\end{figure}

The obtained results of molecular dynamics simulations {have shown} that the dynamics of counterions  in the external electric field is not uniform in different regions of the DNA double helix. The visual examination of the simulation trajectories of the K-DNA system in the electric filed parallel to the double helix axis ($E_\parallel$)  {has revealed} that for the voltage values starting from 870 mV the counterions are hopped by the electric field and move along the macromolecule. This effect is more  visible at higher voltage values (Fig. \ref{fig:snapshots} a). The mobility of the ions in the first hydration shell is governed by the structure of the double helix. The counterions stay longer in the places with narrower minor groove, where they may stuck for longer time. Since the width of the grooves is modulated by the sequence of nucleotide bases the headway of the ion also depends on nucleotide sequence. Due to the thermal fluctuations of the double helix the width of the grooves changes that facilitates the motion of the counterions along DNA macromolecule inside the groove. In the second hydration shell the counterions are free to move and the mobility is about 6 times higher. The exchange of counterions between the different shells is present, but in average the number of counterions is constant. In this regard, the DNA macromolecule may be considered as a wire constructed as a core that makes the counterions condensed.

{The performed} visual examination of K-DNA system in the electric field perpendicular to the helical axis ($E_\perp$) {has shown} the motion of counterions is less affected by the DNA macromolecule that is due to the high {ratio of the volume of the system to the volume of DNA fragment}. Note, in the multi-DNA layers, where the ratio of the system volume to the DNA volume is lower, the conductivity values can be considerably higher in parallel direction to the helical axes rather than in perpendicular ~\cite{aksimentiev2015}. At high voltages the DNA fragment starts to bend  (Fig. \ref{fig:snapshots}b). Within the external bias from 1450 mV to 2030 mV the bend substantially increases (from $\sim$50$^\circ$ to $\sim$90$^\circ$) inducing the breakage of the hydrogen bonds in the complementary nucleotide pairs of the DNA double helix. This effect also manifests itself by the rapid changes in the width of the grooves at high voltages (Fig. \ref{fig:grooves} b). The extension of the minor groove makes DNA permeable for the counterions, {which} is very similar to the increase of the distance between the different layers of DNA origami in electric field~\cite{aksimentiev2015}. Therefore, we can assume that the electric field {induces} the conformational transition of the double helix resembling the melting of the DNA double helix under the action of the external force. The observed threshold value of the external electric field that allows such transition to take place is $\sim$ 2000 mV.

{The  values of ionic conductivities for K-DNA system obtained in our molecular dynamics simulations (the values 3.64 mS/cm and 4.28 mS/cm in Tabl. \ref{tab:results}), are rather close to the experimental value 3.1 mS/cm for Na-DNA solution ~\cite{Kuznetsov1981}. Despite the difference of the potassium and sodium ions, such comparison makes sense because in atomistic molecular dynamics simulations in the present force field  K$^+$ ions reveal the structure-making character of hydration and so do Na$^+$ ions in real solutions ~\cite{perepelytsya2019positively}. The calculated conductivity values are about 4 times lower than the experimental value for the DNA water solution without added salt obtained in the work \cite{Liubysh2014}. However, we should note that the concentration of DNA in our simulations is about two orders of magnitude higher than in the experiments \cite{Kuznetsov1981,Liubysh2014}, while the ionic conductivity is known to decrease with the concentration of DNA ~\cite{Kuznetsov1981}. The comparison of the results for KCl solutions has shown that the value of ionic conductivity obtained in our simulations (the value 8.95 mS/cm in the Tabl. \ref{tab:results}) is close to the experimentally observed value for KCl solution with 0.1 M concentration under 25$^\circ$C  12.82 mS/cm~\cite{KClconductivity1991}. Thus, despite the approximate character of the comparison of our results with the experiments \cite{Kuznetsov1981,Liubysh2014} the calculated conductivity values are in sufficient agreement with the experimental observations.}

The influence of the DNA macromolecule on the dynamics of K$^+$ counterions observed in the present molecular dynamics studies may be interesting for the both understanding the basic properties of the DNA-counterion system and development of the biotechnological applications. In particular, the observed dependence of the ionic current on the sequence of the nucleotide bases (Fig. \ref{fig:grooves} a, top graph) supports the idea of the determination of the sequence of nucleotides in the DNA macromolecule using the information about the ionic current in the system{~\cite{kowalczyk_2012,comer_2012,belkin2016,aksimentiev2015}}. The observed in the present work conformational transition of DNA under the action of the external electric filed (Fig. \ref{fig:snapshots} b) may be {important} for the understanding processes, where the double helix is separated into two strands. In a living cell the opening {of DNA nucleotide pairs is known to take place under the action of enzymes during transcription and reduplication~\cite{singleton2007structure}, but the certain mechanisms of these processes are not studied  sufficiently yet. The opening of DNA base pairs is extensively studied under the temperature influence (DNA melting experiments, see the review~\cite{vologodskii2018dna}) and under the mechanical force (single-molecule manipulation experiments on DNA unzipping~\cite{bockelmann2004dynamics}). In the present research the field-induced melting of the DNA double helix has been observed that may be used for the further studies of physical mechanisms of the opening of DNA base pairs.}

\section{Conclusions}
The molecular dynamics simulations of the DNA double helix with K$^+$ counterions and KCl solution in the external electric field have been carried out. The dynamical response of the ions on the presence of the electric field has been studied. The ionic conductivities and the mobilities of the ions have been calculated for the case of different orientations of the external electric field with respect to the double helix. The results {have shown} that the ions in KCl solution are characterized by the {mobility} values about 0.35 mS/cm. In the case of K-DNA solution the counterions around the double helix are slowed down compared to the free salt solution that is observed by the decrease  of mobility. The counterions that are inside the DNA double helix have the mobility six {times} lower than in the case of free salt solution. In the case of the parallel orientation of the electric field the DNA macromolecule may be considered as a wire consisted of the linear core condensing counterions around. The counterions move along the macromolecule, outside and in the grooves of the double helix, staying longer in the regions with narrower minor groove. Their motion along DNA is modulated by the sequence of nucleotide bases as the width of the grooves of the double helix is sequence-{dependent}. In the case of the electric field perpendicular to the double helix the dynamics of counterions is less affected by DNA, since the time of counterion localization near the double helix is shorter. Under the high voltages (more than $\sim$ 2000 mV) applied {orthogonal to the double helix the conformational transition of the macromolecule has been observed. Thus, the present study has revealed qualitatively different effects of the external electrostatic field applied parallel and orthogonal to the DNA double helix, which may be important for the understanding of counterion distribution around DNA in the real systems}.\\
\\
The authors acknowledge the computational facilities: the BITP grid-cluster and HTCondor system (CHTC UW-Madison resources http://chtc.cs.wisc.edu/ via the access server htcondor.bitp.kiev.ua). {The authors thank Dr. Leonid Belous for his assistance at HTCondor computational facilities.} The present work was partially supported by the Project of the of the National Academy of Sciences of Ukraine (0119U102721). The authors declare no conflict of interest.

\bibliographystyle{spphys} 
\bibliography{literature}   % name your BibTeX data base

\end{document}